\documentclass[epj]{svjour}
\usepackage{amsmath}
\usepackage{bm}
\usepackage{graphicx}
\usepackage{subfig}
\usepackage{amssymb}
\usepackage{color}
\usepackage{cite}
\begin{document}
%%%%%%%%%%%%%%%%%%%%%%%%%%%%%%%%%%%%%%%%%%%%%%%%%%%%%%%%%%%%%%%%%%%%
\title{Optimisation of multifractal analysis using box-size scaling}
\author{Alberto Rodriguez\inst{1,2} \and Louella J. Vasquez\inst{1} \and Rudolf A. R\"omer\inst{1}\mail{r.roemer@warwick.ac.uk}}
\institute{Department of Physics and Centre for Scientific Computing, University of Warwick, 
Coventry, CV4 7AL, United Kingdom \and  Departamento de Fisica Fundamental, Universidad de Salamanca, 37008 Salamanca, Spain}
\date{Received: date / Revised version: \today , $Revision: 1.14 $}
\abstract{We study various box-size scaling techniques to obtain the multifractal properties, in terms of the singularity spectrum $f(\alpha)$, of the critical eigenstates at the metal-insulator transition within the 3-D Anderson model of localisation. The typical and ensemble averaged scaling laws of the generalised inverse participation ratios are considered. In pursuit of a numerical optimisation of the box-scaling technique we discuss  different box-partitioning schemes including cubic and non-cubic boxes, use of periodic boundary conditions to enlarge the system and single and multiple origins  for the partitioning grid are also implemented.  We show that the numerically most reliable method is to divide a system of linear size $L$ equally into cubic boxes of size $l$ for which $L/l$ is an integer. This method is the least numerically expensive while having a good reliability.
 % end of PACS codes
} %end of abstract
%
%\titlerunning{Optimisation of multifractal analysis using box-size scaling}
\authorrunning{A. Rodriguez, L. J. Vasquez and R. A. R\"omer}
\maketitle
\section{Introduction}
\label{sec-intro} 
%%%%%%%%%%%%%%%%%%%%%%%%%%%%%%%%%%%%%%%%%%%%%%%%%%%%%%%%%%%%%%%%%%%%%%%%
The multifractal analysis (MFA) \cite{Jan94a,ChaJ89,MilRS97} of critical electronic wavefunction probabilities $|\psi_i|^2$ in the three-dimensional Anderson model of localisation \cite{And58,LeeR85,AbrALR79,RomS03} is based on the scaling of the generalised inverse participation ratio (gIPR) $P_q$.
For a system with volume $L^d$, the gIPR for normalised wavefunctions is defined as  $P_q (l)\equiv\sum_{k}^{N_l} \mu_k^q(l)$
where the integrated measure $\mu_k(l)=\sum_i^{l^d} |\psi_i|^2$ is computed in all $N_l$  boxes with linear size $l$ covering the system. At criticality the scaling law 
\begin{equation} 
P_q (\lambda)\propto\lambda^{\tau(q)}
\label{eq-scalingIPR}
\end{equation}
is expected to hold in a certain range of values for  $\lambda\equiv l/L$ \cite{Jan94a}. 
The multifractal character is most often studied in terms of the singularity spectrum $f(\alpha)$.
Its true behaviour at criticality can only be found in the thermodynamic limit.  This limit can be reached in \eqref{eq-scalingIPR} by taking the box size to be $l\rightarrow 0$.  In implementing this so-called {\em box-size scaling method}, one usually considers a fixed system size $L$ and partitions it into smaller boxes.
The scaling behaviour of the gIPR  with box size is then obtained by varying $l$ and averaging over many samples.  Hence, with only one system size to be considered, box-size scaling is numerically inexpensive and has been much used previously \cite{MilRS97,SchG91,GruS95}.

\begin{figure*}
	\centering
 	\subfloat[Cubic boxes for integer $L/l$ and $L=6$ and $l=2$]{\includegraphics[width=.2\textwidth]{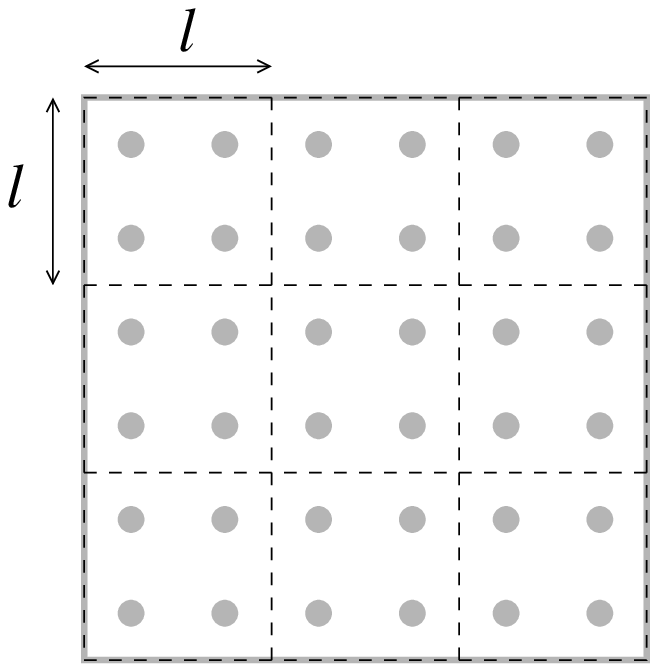}\label{fig-cubicint}}
	\quad
	\subfloat[Unrestricted values of $l$. In this case $L=6$, $l=4$ and for the scaling $\lambda=l/L^\prime$]{\includegraphics[width=.25\textwidth]{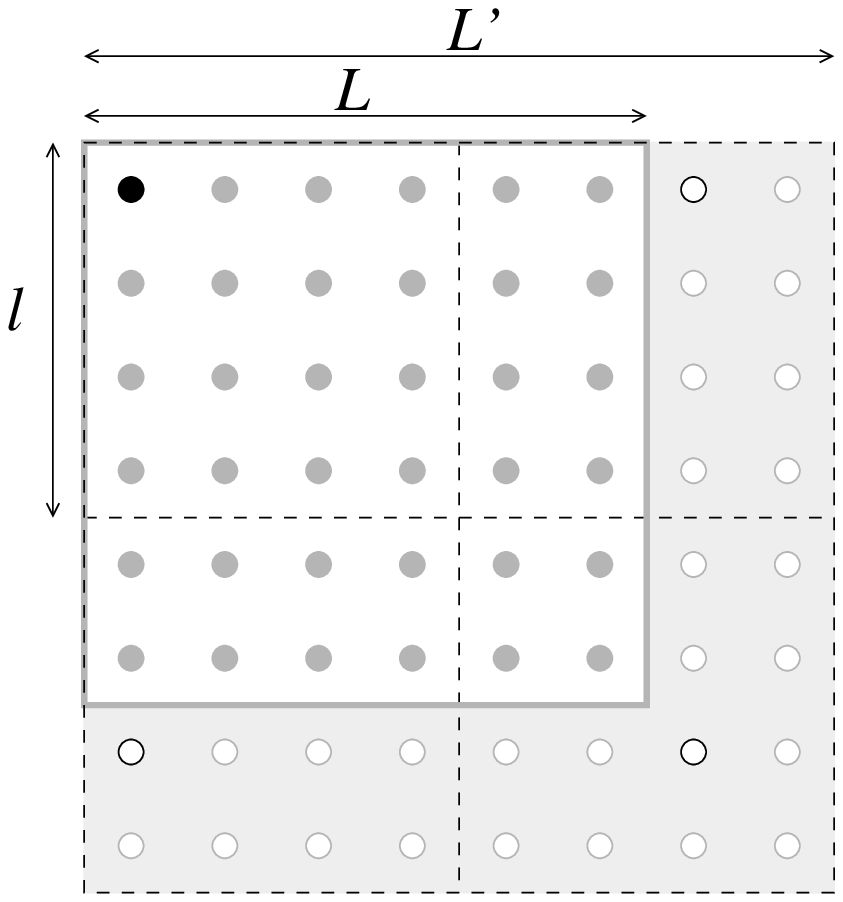}\label{fig-allcubic}}
	\quad
	\subfloat[Non-cubic boxes with $L=6$, $l_x=2$ and $l_y=3$. For the scaling $\lambda=\sqrt{l_x l_y}/L$]{\includegraphics[width=.2\textwidth]{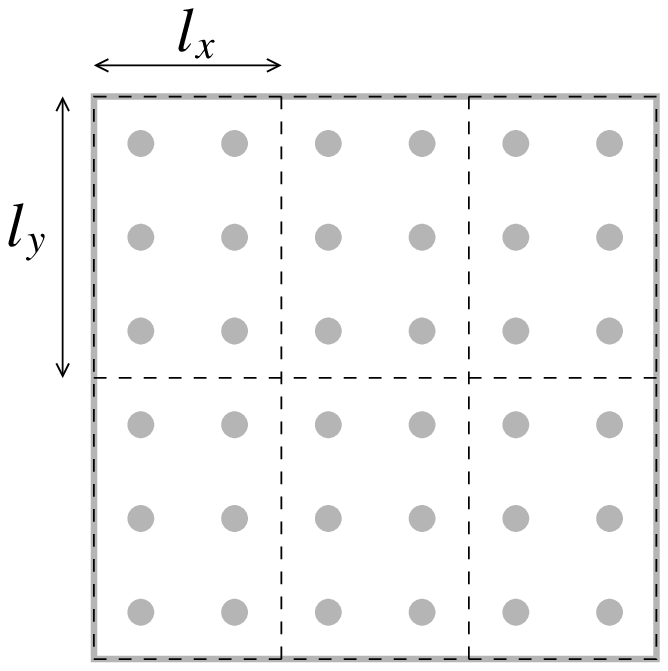}\label{fig-noncubic}}\\
	\subfloat[Multiple origins for cubic boxes and integer $L/l$. The $l^2$ non-equivalent origins for the partitioning are shown for the case $L=6$ and $l=2$]{\includegraphics[width=.7\textwidth]{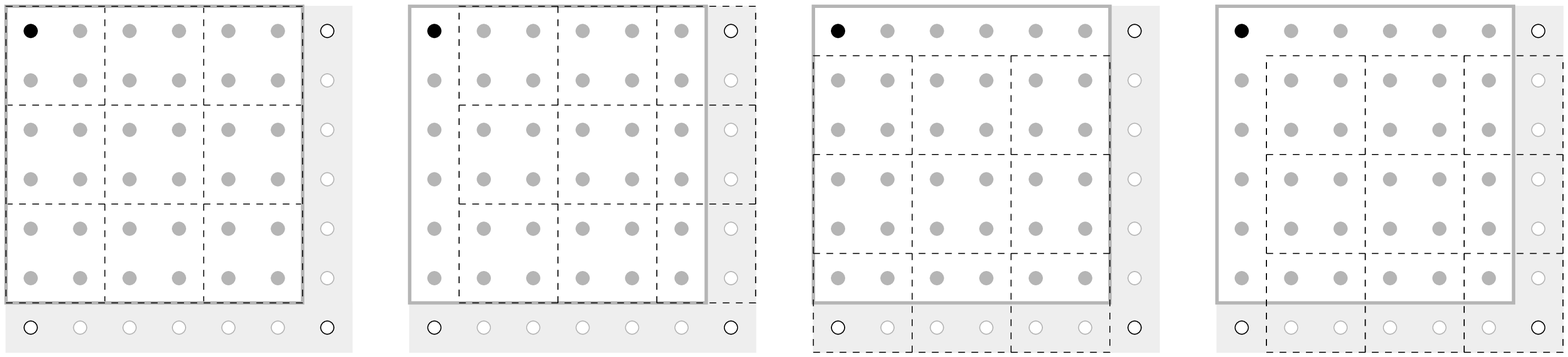}\label{fig-multipleorig}}
 	 \caption{Two-dimensional illustration of the different box-partitioning schemes. The black dashed lines mark the partitioning grid. The thick grey line corresponds to  the boundaries of the physical system with linear size $L$. In cases (b) and (d), the empty points in the shaded regions outside the system are obtained using periodic boundary conditions to properly complete all boxes. The position of the black sites highlights the periodicity pattern.}
	\label{fig-boxpartiotioning} 
\end{figure*}

There exist, however, multiple ways of  carrying out this box partitioning, and some of them might lead to better results, e.g. better fits or an improved statistical analysis. In the present work we analyse the performance of several partitioning schemes, some of which have previously been used in the literature to perform the MFA \cite{SchG91,MilRS97,Cue03}. In particular, we shall study the application of cubic versus non-cubic boxes, the use of periodic boundary conditions to enlarge the system, and multiple origins for the partitioning  grid, and adaptive linear fits. In Fig.\ \ref{fig-boxpartiotioning} we indicate these various strategies schematically.
Here, we show that the use of cubic boxes with integer $L/l$ seems to be best suited to calculate the singularity spectrum using the box-size scaling of $P_q$. 

In Refs.\ \cite{VasRR08a,RodVR08} we had shown that the box-size scaling method is more likely to be affected by finite size effects as compared to the so-called system-size scaling approach. The latter method is based on varying system size $L$ with fixed box size $l$ in the limit $L\rightarrow \infty$ \cite{EveMM08}. Nevertheless, we emphasize that box-size scaling  is still a useful alternative since (i) the computational and data-storage requirements are much less demanding than for system-size scaling, (ii) in certain regions of the spectrum such as close to the maximum, box-size scaling is also quite accurate, and (iii) in some situations it might be the only applicable method, i.e.\ when carrying out an MFA of experimental data \cite{LebL08,MorKMG02,HasSMI08} for which the system size cannot be easily changed. It is therefore important to know how to maximise the use and performance of box-size scaling.

%%%%%%%%%%%%%%%%%%%%%%%%%%%%%%%%%%%%%%%%%%%%%%%%%%%%%%%%%%%%%%%%%%%%%%%%
\section{The Anderson model of localisation and its numerical diagonalisation}
\label{sec-model}
%%%%%%%%%%%%%%%%%%%%%%%%%%%%%%%%%%%%%%%%%%%%%%%%%%%%%%%%%%%%%%%%%%%%%%%%
We use the tight-binding Anderson Hamiltonian in lattice site basis as given by
\begin{equation} \label{anderson_H1} 
\mathcal{H}=\sum_{i} \varepsilon_i~\vert i\rangle\langle i\vert + \sum_{i\neq j} t_{ij}~\vert i\rangle\langle j\vert,
\end{equation}
where site $i=(x,y,z)$ is the position of an electron in a cubic lattice of volume $V=L^3$, $t_{ij}$ are nearest-neighbour hopping amplitudes and 
$\varepsilon_i$ is the random site potential energy.
The hopping amplitude is taken to be $t=1$  and periodic boundary conditions (PBC) are used to minimise boundary effects.
We consider $\varepsilon_i$ to have a box probability distribution in the interval $[-W_c/2,~W_c/2]$, where $W_c$ is taken to be the strength of the critical value of the disorder.
We assume $W_c=16.5$, above which all eigenstates are localised \cite{SleMO03,SleMO01,OhtSK99,MilRSU00}.
We have considered eigenstates $\psi=\sum_{i} \psi_i |i\rangle$ only in the vicinity of the band centre $E=0$.  Moreover, we only take about five eigenstates in a small energy window at $E=0$ for any given realization of disorder \cite{VasRR08a,RodVR08}.  Since the aim of the present study is to compare various box-partitioning schemes, we shall restrict ourselves to moderate system sizes of $L^3=60^3= 216000$. The resulting sparse matrices have been diagonalised using JADAMILU \cite{BolN07}.

%%%%%%%%%%%%%%%%%%%%%%%%%%%%%%%%%%%%%%%%%%%%%%%%%%%%%%%%%%%%%%%%%%%%%%%%
% BRIEF THEORY
%%%%%%%%%%%%%%%%%%%%%%%%%%%%%%%%%%%%%%%%%%%%%%%%%%%%%%%%%%%%%%%%%%%%%%%%
\section{Typical and ensemble averaging for the MFA} \label{sec-mfa}
The well known singularity spectrum $f(\alpha)$ is defined from the $\tau(q)$ exponents via a Legendre transformation: $f_q \equiv f(\alpha_q)=q\alpha_q-\tau(q)$ and $\alpha_q=\tau^\prime(q)$.  The physical meaning of the $f(\alpha)$ is as follows.  It is the fractal dimension of the set of points where the wavefunction intensity is
$|\psi_i|^2\sim L^{-\alpha}$, that is in a discrete system the number of such points $N_\alpha$ scales as $L^{f(\alpha)}$.
The numerical multifractal analysis is based on an \textsl{averaged} form of the scaling law \eqref{eq-scalingIPR} for the gIPR in the limit $\lambda\equiv{l}/{L}\rightarrow 0$, where the contributions from several finite-size critical wavefunctions corresponding to different disorder realizations are properly taken into account.
In the following, we give the typical- and ensemble-averaged form of the singularity strength $\alpha$ and singularity spectrum $f(\alpha)$.  The expressions are written to be optimal for direct numerical computation. Their compact forms are given in Refs.\  \cite{VasRR08a,RodVR08}.
%%%%%%%%%%%%%%%%%
%\subsection{Typical Averaging}
The scaling law for the {\em typical} average of the moments $P_q$ is defined as
\begin{equation}
 	e^{\left<\ln P_q(\lambda)\right>}\propto\lambda^{\tau^{\textrm{typ}}(q)},
 	\label{eq-TYPscale}
\end{equation}
where $\langle\cdots\rangle$ denotes the arithmetic average over many realizations of disorder, i.e.\ over many different wavefunctions at criticality. 
The scaling exponents are then defined by
\begin{equation}
 \tau^\textrm{typ}(q) = \lim_{\lambda\rightarrow 0}\frac{\langle \ln P_q (\lambda)\rangle}{\ln \lambda}.%
\label{eq-tautyp}
\end{equation}
Applying a Legendre transformation \cite{Jan94a}, we obtain the definitions for $\alpha^\textrm{typ}_q$ and $f^\textrm{typ}_q$,
\begin{subequations}
\begin{align}
\alpha^\textrm{typ}_q =  \lim_{\lambda\rightarrow0}\frac{1}{\ln\lambda} &\left\langle \frac{1}{P_q(\lambda)}\sum_{k=1}^{N_\lambda}\mu_k^q(\lambda)\ln\mu_k(\lambda) \right\rangle \label{eq-aqtyp} \\
f(\alpha^\textrm{typ}_q)= \lim_{\lambda\rightarrow0}\frac{1}{\ln\lambda} & \bigg[q\left< \frac{1}{P_q(\lambda)}\sum_{k=1}^{N_\lambda}\mu_k^q (\lambda)\ln\mu_k(\lambda)\right> \notag 
\\ &\quad -\left<\ln P_q(\lambda)\right>\bigg].
\label{eq-fqtyp}
\end{align}
\label{eq-falfatyp}
\end{subequations}
%%%%%%%%%%%%%%%%
%\subsection{Ensemble Averaging}
The scaling law for the {\em ensemble} average involves the arithmetic average of $P_q$ over all realizations of disorder,
\begin{equation}
 	\left<P_q (\lambda) \right>\propto\lambda^{\tau^\textrm{ens}(q)}.
	\label{eq-ENSscale}
\end{equation}
Thus the definition of the scaling exponents is 
\begin{equation}
 	\tau^\textrm{ens} (q) = \lim_{\lambda\rightarrow 0}\frac{\ln \langle P_q (\lambda)\rangle}{\ln \lambda},
	\label{eq-tauens}
\end{equation}
and the corresponding definitions of $\alpha^\textrm{ens}_q$ and $f^\textrm{ens}_q$ can be written as
\begin{subequations}
\begin{align}
\alpha^\textrm{ens}_q =  \lim_{\lambda\rightarrow0} \frac{1}{\ln\lambda} & \frac{1}{\left<P_q(\lambda)\right>}\left\langle \sum_{k=1}^{N_\lambda}\mu_k^q(\lambda)\ln\mu_k(\lambda) \right\rangle \\
 f(\alpha^\textrm{ens}_q) = \lim_{\lambda\rightarrow0} \frac{1}{\ln\lambda} &\bigg[\frac{q}{\left<P_q(\lambda)\right>}\left< 
\sum_{k=1}^{N_\lambda}\mu_k^q(\lambda)\ln\mu_k(\lambda)\right> \notag\\ & \quad -\ln\left<P_q(\lambda)\right>\bigg].
\end{align}
\label{eq-falfaens}
\end{subequations}
The values of $\alpha$ and $f(\alpha)$ are obtained from the slopes of the linear fits of the averaged contributions in Eqs.~\eqref{eq-falfatyp} and \eqref{eq-falfaens} (typical and ensemble respectively) versus $\ln \lambda$, which is calculated for different values of the box-size $l$. Both averages in the thermodynamic limit are expected to tend towards the same spectrum for $f(\alpha)>0$. We refer the reader to Refs.\ \cite{MirE00,EveMM08,RodVR08} for a more detailed discussion of the relationship between both averages as used in an MFA. 
%%%%%%%%%%%%%%%%%%%%%%%%%%%%%%%%%%%%%%%%%%%%%%%%%%%%%%%%%%%%%%%%%%%%%%%%
% DISCUSSION
%%%%%%%%%%%%%%%%%%%%%%%%%%%%%%%%%%%%%%%%%%%%%%%%%%%%%%%%%%%%%%%%%%%%%%%%
%\section{Partitioning Schemes}
%\label{sec-partition}
%%%%%%%%%%%%%%%
\section{Partitioning into cubic boxes with integer ratios $L/l$}
\label{ssec-cubicinteger}
\begin{figure}
  \centering
   \includegraphics[width=.95\columnwidth]{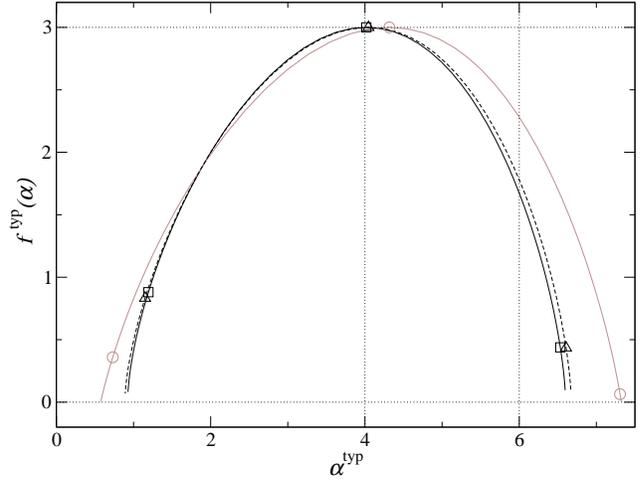}
  \caption{Singularity spectrum obtained from box-size scaling of the typical average of $P_q$ for $L=60$ averaging over $10^3$ states using different schemes of box-partitioning: cubic boxes for integer $L/l$ (solid black), cubic boxes for all values of $l$ (grey), non-cubic boxes (dashed black).  The symbols denote the position of $q=2$ (left), $q=0$ (centre) and $q=-4$ (right).}
\label{fig-typ}
\end{figure}

The most simple way of partitioning  the system with linear size $L$ is to use an isotropic cubic grid which can fit exactly the system.
In this case the box-sizes in each direction satisfy $l_x=l_y=l_z$ with values of $l$ such that $L/l\equiv n \in \mathbb{N}$, and thus we always cover the system of size $L^3$ with an integer number of boxes, as illustrated in Fig.~\ref{fig-cubicint}.  The singularity spectrum $f(\alpha)$ obtained using this partitioning method for a system with $L=60$ after taking the typical average over $10^3$ states is shown in Fig.~\ref{fig-typ}.
It must be emphasized that since we are dealing with a discrete system, the values of the box-size must be larger than the lattice spacing, in order  to observe properly the multifractal fluctuations of the distribution \cite{Jan94a,VasRR08a}. 
Usually we consider values for the box-size  in the interval $l\in[10, L/2]$. The points used for the linear fits shown in Fig.\ \ref{fig-typ-fits} correspond to $l= 10, 12, 15, 20, 30$. 
Although easy to implement numerically, the drawback of  the present  method is that depending on the system size sometimes only a few values for $l$ are allowed and that imposes a restriction on the reliability of the fits to obtain $\alpha_q$ and $f_q$. In the following sections we consider several additional partitioning strategies and compare their performance to the basic integer cubic-boxes technique.
%%%%%%%%%%%%%
\section{Using cubic boxes with unrestricted values for $l$}
\label{ssec-allvalues}
\begin{figure}
  \centering
  \includegraphics[width=.95\columnwidth]{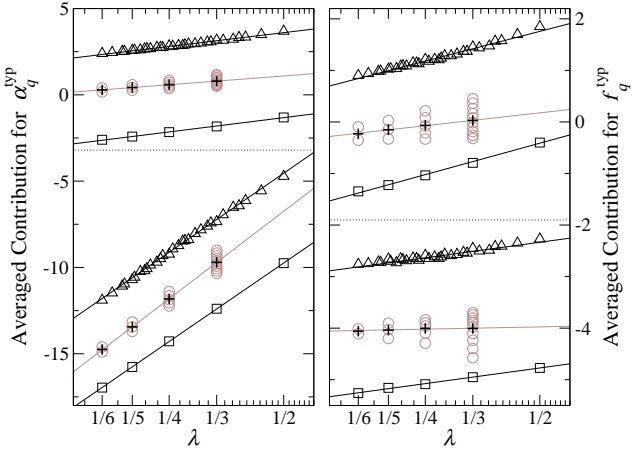}
  \caption{Linear fits of Eqs.~\eqref{eq-falfatyp} for $\alpha^\textrm{typ}_q$ (left) and $f^\textrm{typ}_q$ (right) values of the singularity spectra shown in Fig.~\ref{fig-typ} using different schemes of box-partitioning: cubic boxes for integer $L/l$ (black, square), cubic boxes for all values of $l$ (grey, circle), non-cubic boxes (black, triangle).  For the method of cubic boxes for all values of $l$, the data point corresponding to the average contribution for each $\lambda$ (plus symbol) is also shown.
The dotted lines separate fits for $q=2$ (top) from $q=-4$ (bottom). The values of $\alpha^\textrm{typ}_q$ and
$f^\textrm{typ}_q$ are given by the slopes of the fits. Data points have been properly shifted vertically to ensure optimal visualisation.  In all cases, the standard deviations are contained within symbol size.}
\label{fig-typ-fits}
\end{figure}

As illustrated in panel (b) of Fig.~\ref{fig-boxpartiotioning}, an alternative partitioning method would be to use \textit{all} values of $l$ in the range $10\leqslant l<L/2$.
The system is partitioned into cubic boxes without imposing the restriction of $L/l$ being an integer.  As such, there are values of $l$ where some of the outer boxes will have a lack of sites. This can be overcome by using periodic boundary conditions where all necessary values $|\psi_i|^2$ will be repeated until all boxes are properly filled. One must realise that although the diagonalisation process to obtain the eigenstates uses PBC to minimise edge effects, no true periodicity pattern exists in the wavefunction itself. However by using this partitioning scheme we are imposing a periodicity pattern on the distribution of $|\psi_i|^2$ that in principle might distort its multifractal properties.
Furthermore, we must emphasize here that repeating $|\psi_i|^2$ as seen in Fig.~\ref{fig-allcubic} will mean enlarging the system into $L'$ and hence a renormalisation of the wavefunctions\footnotemark[1] must be carried out if one wants to use the equations in Section \ref{sec-mfa}.
 \footnotetext[1]{Alternatively the scaling law for the gIPR, as well as the definitions of $\tau(q)$, $\alpha_q$ and $f(\alpha_q)$ can be generalised to account for unnormalised distributions.}
In Fig.~\ref{fig-typ}, a comparison can be found between the typical singularity spectra obtained using the current box-partitioning method and the one involving cubic-boxes with integer $L/l$. The corresponding linear fits used to obtain  $\alpha^\textrm{typ}_q$ and $f^\textrm{typ}_q$ for a couple of values of $q$ are shown in Fig.~\ref{fig-typ-fits}.

Using this partitioning approach the number of available $\lambda$ values in the fits is not increased, since whenever the system is enlarged the correct value for the scaling variable must be also  redefined to $\lambda=l/L^\prime$, which is the inverse of the number of boxes needed to cover the system in each dimesion. Therefore different box-sizes give contributions to the same $\lambda$ value in the fit. All these contributions as well as the corresponding weighted average are explicitly shown in Fig.~\ref{fig-typ-fits}. The linear fits performed to obtain the singularity spectrum are based on a $\chi^2$ minimisation taking into account the individual standard deviations of each point. The first thing to notice is that the uncertainties of the individual contributions to a given $\lambda$  are clearly in conflict with the range of values spanned by the points. In fact we have observed that there exists a defined tendency of behaviour for these contributions. In the fits for $\alpha_q$ (left panel in Fig.~\ref{fig-typ-fits}), for large positive $q$ and a fixed value of $\lambda$ the higher contribution is given by the box-size $l$ which does not require an extension of the system. As $l$ is increased and the system size is enlarged, preserving the value of $\lambda$, the contributions are progressively smaller. This makes the value of the slope, that is $\alpha_q$, go systematically to smaller values at the end of the left branch of the spectrum, when compared to the method in Sec.~\ref{ssec-cubicinteger}. For high negative $q$ the behaviour is the opposite, for a given $\lambda$ the additional contributions of the box-sizes requiring system enlargement grow with the box-size, and therefore the slope attains higher values, i.e.  $\alpha_q$  moves towards higher values at the end of the right branch of the spectrum. This leads to a systematic broadening of $f(\alpha)$ as shown in Fig.~\ref{fig-typ} . The effect of the additional contributions in the fits for $f_q$ (right panel in Fig.~\ref{fig-typ-fits}) causes a decrease of the slopes for high $q$, either negative or positive.
 We have also checked that the shape of the singularity spectrum is strongly dependent on the range of box-sizes taken into account for the linear fit. All these facts clearly make the present partitioning method very unreliable and not suitable to perform a numerical  MFA. 
%%%%%%%%%%
\section{Partitioning with rectangular boxes}
\label{ssec-noncubic}
Another strategy is to consider an anisotropic non-cubic box partitioning as shown in Fig.~\ref{fig-noncubic}, with at least one of the linear sizes $l_x,l_y,l_z$ different from the other two. 
For simplicity we will consider values for the box sides such that $L/l_{x,y,z}$ is an integer. Hence in all directions the system can be covered exactly with an integer number of boxes, although this number can be different for each direction $x,y,z$. The scaling parameter used in the present case is $\lambda=\bar{l}/L$ with an effective linear box-size defined as $\bar{l}=(l_x l_y l_z)^{1/3}$. Let us note that different combinations of $\{l_x,l_y,l_z\}$ can lead to the same $\bar{l}$ and thus we could have different contributions to the same $\lambda$, as in the method described in Sec.~\ref{ssec-allvalues}.  In  the fits of Fig.~\ref{fig-typ-fits} only the averaged contribution for each $\lambda$ is shown.
The corresponding singularity spectrum (Fig.~\ref{fig-typ}) is similar to the $f(\alpha)$ obtained using only cubic boxes with integer $L/l$. We again note a tendency to broaden but it is less pronounced than when using unrestricted values for the box-size. 
Although in the present case the number of points used in the linear fits is noticeably increased as seen in Fig.~\ref{fig-typ-fits}, 
no significant improvement with regards to the reliability of the $f(\alpha)$ has been observed.

%%%%%%%%%%
\section{Using multiple origins for the box partitions}
\label{ssec-multipleorig}
\begin{figure}
  \centering
  \includegraphics[width=.95\columnwidth]{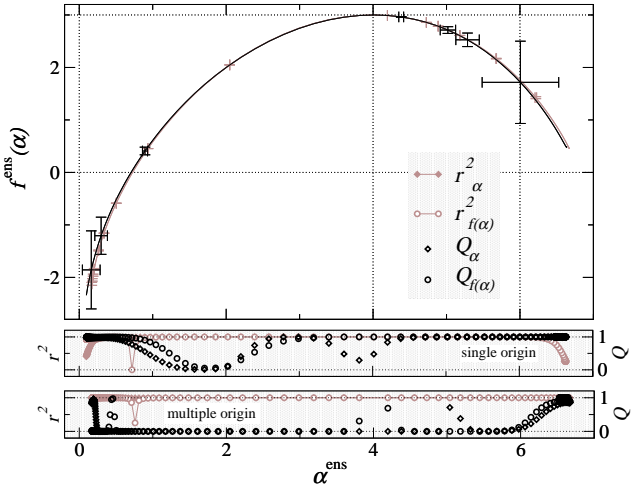}
  \caption{Singularity spectrum obtained from box-size scaling for $L=60$ averaging over $10^3$ states. Ensemble average using cubic boxes for integer $L/l$ with single-origin (black) and multiple origins (grey) for the box partitioning.
The error bars are equal to one standard deviation.  The values for the linear correlation coefficient $r^2$ and quality-of-fit parameter $Q$ of the fits to obtain $\alpha^\textrm{ens}_q$ and $f^\textrm{ens}_q$ in the single origin and multiple origin cases are shown in the bottom shaded panels as labelled.}
\label{fig-ens}
\end{figure}

An further possibility to increase the reliability of the multifractal spectrum would be to consider different origins for the partitioning method as demonstrated in Fig.~\ref{fig-multipleorig}. This means that instead of considering the box-partitioning from a single origin, we could use different points in the system as origins.  This is equivalent, once we have done the partitioning for a given box-size $l$ and we have a rigid partition grid, to shift the whole wavefunction at the same time one site at a time in all directions a maximum of $l$ sites. Here, one uses the PBC in such a way that the wavefunction sites that are left of the grid in one direction enter the system through the opposite end. Note that each value of $|\psi_i|^2$ is used only once. Hence for each box-size we have $l^3$ different non-equivalent origins for the box-partitioning. Therefore for a given box-size $l$ the number of contributions in the average of the gIPR $P_q$ is multiplied by $l^3$.  Since the ensemble average is specially sensitive to the number of disorder realizations considered, we show in Fig.~\ref{fig-ens} how $f^\textrm{ens}(\alpha)$  behaves when multiple origins are taken into account. 
The change with respect to the single-origin situation \cite{RodVR08} is really insignificant, there is almost a perfect overlap for both lines. But the standard deviations are reduced almost to zero when multiple origins are considered. This is due to the fact that the standard deviations of the averaged contributions for the points in the fits, include a term $1/\sqrt{\mathcal{N}}$ where $\mathcal{N}$ is the number of states we average over. Since every origin counts as a different contribution, then the uncertainties are greatly reduced. However, let us show that this reduction is in fact misleading and does not really mean an increase in reliability. In the bottom panels of Fig.~\ref{fig-ens}, we compare the quality-of-fit parameter $Q$ of the linear fits to obtain $\alpha$ and $f(\alpha)$ for the single-origin and multiple-origin techniques. In both cases the linear correlation coefficient $r^2$ is close to $1$ in the whole $\alpha$ range indicating a good linear behaviour. But in the multiple-origin situation, the values of $Q_\alpha$ and $Q_{f(\alpha)}$ decrease considerably suggesting that the uncertainties of the points in the fits have been clearly underestimated. Therefore there is no gain in reliability. Considering multiple origins maybe seen as considering eigenstates of other disordered systems which correspond to cyclic permutations in the three spatial directions of the initial disorder realization. Clearly, such transformations will not give independent disorder realizations. In fact the contribution of all these wavefunctions related by cyclic permutations must be very similar, since their probability distribution function for $|\psi_i^2|$ values is identical. Therefore it is not correct to consider their contribution as a different disorder realization. That explains why the spectrum hardly changes when compared  to the single-origin situation. 
From our point of view this strategy should not be considered as a good method to reduce the uncertainty. It must also be emphasized that the multiple-origin strategy is very expensive in terms of computational time. 

%%%%%%%%
\section{An adaptive MFA-fit strategy}
\label{sec-otherapproach}
Aside from the different schemes of partitioning, let us also study another strategy related to the selection of values of box-sizes that are considered for the linear fit. Having in mind that the proper region of values for $l$ to do the scaling might also depend on the value of $q$, we implemented an adaptive-linear-fit strategy. For a given value of $q$, we consider a window with a certain number of points to do the fit and  we maximise the values of $r^2$ and $Q$ by shifting this window throughout all the available interval of values for $l$, only for  box-sizes with integer $L/l$. In this way for each $q$ the best linear behaviour is achieved using a different region of contiguous $l$ values.  In this case we have considered a system with size $L=210$, for which $14$ available values of $l$ exist in the interval $l\in[2,L/2]$, and we tried different widths of the window for $l$. Unfortunately using this strategy we have not been able to see proper multifractal spectra. Due to the fluctuations of the uncertainty for the points in the fits, sometimes the shifting of the window of $l$-values does not follow a smooth tendency with $q$, moreover it also happens that for a given $q$ the best fit for $\alpha_q$ and $f_q$ is achieved in a different range of $l$. These effects give rise to the appearance of discontinuities and irregularities in $f(\alpha)$.  Apparently our initial premise is not true. 
%%%%%%%%%%%%%%%%%%%%%%%%%%%%%%%%
%%%%%%%%%%%%%%%%%%%%%%%%%%%%%%%%
\section{Conclusion}
We have shown that the simple box-partitioning MFA method based on the use of cubic boxes with integer side length $L/l$ and a single-origin for the partitioning, although bearing some limitations, is numerically the most reliable one. It is also optimal in terms of time and computational requirements, and it has already been successfully implemented to obtain the scaling behaviour at the Anderson transition for very large system sizes \cite{VasRR08a,RodVR08}. Therefore it should be considered the method of choice.
Furthermore, let us point out that recently \cite{MirFME06} an analytical symmetry relationship for $f(\alpha)$ in the thermodynamic limit has been given. This relation states that the spectrum must be contained within $0\leqslant \alpha\leqslant 6$ for bulk systems \cite{ObuSFGL07,ObuSFG08}. Our results in Fig.~\ref{fig-typ} show that the strategy of cubic boxes with integer sizes $L/l$ and the use of a single origin indeed gives the best agreement with the symmetry \cite{VasRR08a,RodVR08}. 

%%%%%%%%%%%%%%%%%%%%%%%%%%%%%%%%%%%%%%%%%%%%
%%%%%%%%%%%%%%%%%%%%%%%%%%%%%%%%%%%%%%%%%%%
\section*{Acknowledgements}
LJV and RAR gratefully acknowledge EPSRC (EP/C007042/1) for financial support. AR acknowledges financial support from the Spanish government under contracts JC2007-00303, FIS2006-00716 and MMA-A106/2007, and JCyL under contract SA052A07.
%%%%%%%%%%%%%%%%%%%%%%%%%%%%%%%%%%%%%%%%%%%%%%%%%%%%%%%%%%%%%%%%%%%%%%%%
%%%%%%%%%%%%%%%%%%%%%%%%%%%%%%%%%%%%%%%%%%%%%%%%%%%%%%%%%%%%%%%%%%%%%%%%
%
% References
%
%%%%%%%%%%%%%%%%%%%%%%%%%%%%%%%%%%%%%%%%%%%%%%%%%%%%%%%%%%%%%%%%%%%%%%%%


\begin{thebibliography}{99}

\bibitem{Jan94a}
M. Janssen, Int. J. Mod. Phys. B {\bf 8},  943  (1994).

\bibitem{ChaJ89}
A.~B. Chabra and R.~V. Jensen, Phys. Rev. Lett. {\bf 62},  1327  (1989).

\bibitem{MilRS97}
F. Milde, R.~A. {R\"{o}mer}, and M. Schreiber, Phys. Rev. B {\bf 55},  9463
  (1997).

\bibitem{And58}
P.~W. Anderson, Phys. Rev. {\bf 109},  1492  (1958).

\bibitem{LeeR85}
P.~A. Lee and T.~V. Ramakrishnan, Rev. Mod. Phys. {\bf 57},  287  (1985).

\bibitem{AbrALR79}
E. Abrahams, P.~W. Anderson, D.~C. Licciardello, and T.~V. Ramakrishnan, Phys.
  Rev. Lett. {\bf 42},  673  (1979).

\bibitem{RomS03}
R.~A. {R\"{o}mer} and M. Schreiber,  in {\em The {Anderson} Transition and its
  Ramifications --- Localisation, Quantum Interference, and Interactions},
  Vol.~630 of {\em Lecture Notes in Physics}, edited by T. Brandes and S.
  Kettemann (Springer, Berlin, 2003), Chap.~Numerical investigations of scaling
  at the {Anderson} transition, pp.\ 3--19.

\bibitem{SchG91}
M. Schreiber and H. Grussbach, Phys. Rev. Lett. {\bf 67},  607  (1991).

\bibitem{GruS95}
H. Grussbach and M. Schreiber, Phys. Rev. B {\bf 57},  663  (1995).

\bibitem{Cue03}
E. Cuevas, Phys. Rev. B {\bf 68},  024206  (2003).

\bibitem{VasRR08a}
L.~J. Vasquez, A. Rodriguez and R.~A. {R\"{o}mer}, Phys. Rev. B (2008), submitted, {ArXiv}:cond-mat/0807.2217v1.

\bibitem{RodVR08}
A. Rodriguez, L.~J. Vasquez and R.~A. {R\"{o}mer}, Phys. Rev. B (2008), submitted, {ArXiv}:cond-mat/0807.2209v1.

\bibitem{EveMM08}
F. Evers, A. Mildenberg, and A.~D. Mirlin, Phys. Stat. Sol. b {\bf 245},  284  (2008).

\bibitem{LebL08}
M.~A. Lebyodkin and T.~A. Lebedkina, Phys. Rev. E {\bf 77},  026111  (2008).

\bibitem{MorKMG02}
M. Morgenstern {\it et~al.}, Phys. Rev. Lett. {\bf 89},  136806  (2002),
  {ArXiv}: cond-mat/0202239.

\bibitem{HasSMI08}
K. Hashimoto {\it et~al.}, Phys. Rev. Lett. (2008), submitted,
  {ArXiv}: cond-mat/0807.3784

\bibitem{SleMO03}
K. Slevin, P. Marko\u{s}, and T. Ohtsuki, Phys. Rev. B {\bf 67},  155106
  (2003).

\bibitem{SleMO01}
K. Slevin, P. Marko\u{s}, and T. Ohtsuki, Phys. Rev. Lett. {\bf 86},  3594
  (2001).

\bibitem{OhtSK99}
T. Ohtsuki, K. Slevin, and T. Kawarabayashi, {Ann. Phys. (Leipzig)} {\bf 8},
  655  (1999), {ArXiv}: cond-mat/9911213.

\bibitem{MilRSU00}
F. Milde, R.~A. {R\"{o}mer}, M. Schreiber, and V. Uski, Eur. Phys. J. B {\bf
  15},  685  (2000), {ArXiv}: cond-mat/9911029.

\bibitem{BolN07}
M. Bollh\"{o}fer and Y. Notay, Comp. Phys. Comm. {\bf 177},  951  (2007).

\bibitem{MirE00}
A.~D. Mirlin and F. Evers, Phys. Rev. B {\bf 62},  7920  (2000).

\bibitem{MirFME06}
A.~D. Mirlin, Y.~V. Fyodorov, A. Mildenberg, and F. Evers, Phys. Rev. Lett.
  {\bf 97},  046803  (2006).

\bibitem{ObuSFGL07}
H. Obuse {\it et~al.}, Phys. Rev. Lett. {\bf 98},  156802  (2007).

\bibitem{ObuSFG08}
H. Obuse {\it et~al.}, Physica E {\bf 40},  1404  (2008).
\end{thebibliography}
\end{document}